\documentclass{svmult}
  
\usepackage{mathptmx,hyperref}   
\usepackage{helvet} 
\usepackage{courier}
\usepackage{amsmath,amsfonts}
\usepackage{graphicx}
\usepackage[bottom]{footmisc}          
    
\usepackage{mathptmx}         
\usepackage{helvet}            
\usepackage{courier}            
\usepackage{type1cm}             
\usepackage{makeidx}             
\usepackage{graphicx}          
\usepackage{multicol}          
\usepackage[bottom]{footmisc}  
     
\usepackage{epsfig}     
\usepackage{psfrag,rotating}     
\usepackage{amssymb,floatflt,enumerate} 
\usepackage{amsmath,amscd,psfrag,leqno} 
\usepackage[mathscr]{eucal}  
\usepackage{shadethm}           
\usepackage{graphicx,subfigure}   
\usepackage{overpic,contour}       
\contourlength{0.3mm}

\def\Vkt#1{{\mathbf #1}}

\makeindex

\begin{document}

\title*{ 
From axial C-hedra to general P-nets}

\author{G. Nawratil}
\authorrunning{G. Nawratil}
\institute{
  Institute of Discrete Mathematics and Geometry \&  
	Center for Geometry and Computational Design, TU Wien,
  \email{nawratil@geometrie.tuwien.ac.at}}

%
%
\maketitle

\abstract{
We give a full classification of continuous flexible discrete axial cone-nets, which are called axial C-hedra. 
The obtained result can also be used to construct their semi-discrete analogs.   
Moreover, we identify a novel subclass within the determined class of (semi-)discrete axial cone-nets, whose members are named axial P-nets as they fulfill the proportion (P) of the intercept theorem. Known special cases of these axial P-nets are the smooth and discrete conic crease patterns with reflecting rule lines. 
By using a parallelism operation one can even generalize axial P-nets. The resulting general  P-nets  constitute a rich novel class of continuous flexible  (semi-)discrete surfaces, which allow direct access to their spatial shapes by three control polylines. 
This intuitive method makes them suitable for transformable design tasks using interactive tools.
}


\section{Introduction}\label{sec:intro}

A planar quad-surface (PQ-surface) is a plate-and-hinge structure made of quadrilateral panels connected by rotational (R) joints in the combinatorics of a square grid.  
Such a surface is called {\it continuous flexible} (or {\it rigid-foldable} or {\it isometric deformable}) if it can be continuously transformed by a change of the dihedral angles only. In general a rigid-foldable PQ-surface has one degree of freedom (DoF), thus the change of its shape can be controlled by a single active R-joint. 

It is well known \cite[Theorem 3.2]{schief} that PQ-surfaces have a continuous flexion if and only if every $(3\times 3)$ substructure is isometrically deformable. 
Based on spherical kinematic geometry \cite{stachel}, a partial classification of rigid-foldable  $(3\times 3)$ building blocks was obtained by Stachel and the author \cite{NS10,Naw11,Naw12}. 
Inspired by this approach, Izmestiev \cite{Izm17} obtained a full classification of continuous flexible $(3\times 3)$ complexes. 

Note that the rigid-foldability of PQ-surfaces
is not a property of the extrinsic geometry
but of the intrinsic one, which is determined by the
corner angles of the planar quads. 
Nonetheless, certain classes of rigid-foldable 
PQ-surfaces, so-called V-hedra and T-hedra, allow for direct access to their spatial shape through the use of control polylines. 
Due to this intuitive design methods these classes, which were originally introduced by 
Sauer and Graf \cite{graf} in 1931, recently attracted attention in the context of transformable design (see \cite{voss,KMN1,SNRT2021} and the references therein). 
These T-hedra and V-hedra (and their related surfaces \cite{voss}) are more or less\footnote{Beside them only some initial stitching solutions of different $(3\times 3)$ building blocks were presented in the literature \cite{dieleman,he}.} the only known flexible PQ-surfaces
which go beyond the rather abstract classification of $(3\times 3)$  building blocks \cite{Izm17}. 

For the paper at hand the {\it V-hedra}, which are discrete analogs of Voss\footnote{The letter V  stands for Voss in the nomenclature V-hedron.} surfaces \cite{voss_orig},  
are not of interest, in contrast to T-hedra. These discrete analogs of {\it profile-affine surfaces} \cite{graf,sauer}  are briefly recapped at the begin of the next section.
\begin{remark}
Jiang et al.\ \cite{jiang} presented  an optimization technique for an approximate  penalization  of isometrically deformed surfaces with planar quads. The method's design space is restricted to PQ-surfaces which can be seen as discretizations of continuous flexible smooth surfaces (e.g.\ Voss surfaces, profile-affine surfaces).   \hfill $\diamond$
\end{remark}

\subsection{Motivation, Outline and Review}\label{sec:MOR}

A {\it T-hedron} (also known as {\it discrete T-surface}) can be considered as a generalization of a discrete surface of revolution in such a way that the vertically aligned axis of rotation is not fixed but traces the so-called {\it prism polyline} on the base plane $\tau_0$, which is orthogonal to the axis direction. Moreover, the action does not need to be a pure rotation but can be combined with an axial dilatation. By applying this discrete kinematic generation iteratively to  a {\it profile polyline} located in plane $\pi_0$ through the initial axis $q_0$, a quad-surface with planar trapezoidal\footnote{The letter T  stands for trapezoidal in the nomenclature T-hedron.} faces is obtained. 
The polygonal path of the profile vertex located in the base plane is called {\it trajectory polyline}. Therefore, the complete geometric information of the T-hedron is encoded within three polylines (prism, profile, trajectory). 

If only the profile polyline is replaced by a smooth profile curve, we end up with a so-called semi-discrete\footnote{This is the semi-discretization of the vertical kind \cite{KMN1}. There is also a second one of the horizontal kind, but in the paper at hand the semi-discretization always refers to the vertical kind.} T-surface. 

Every vertical strips of a discrete (semi-discrete) T-surface is a patch of a discrete (smooth) cylinder due to its construction. From the projective point of view these cylinders are cones with their tips located on the ideal line of the base plane $\tau_0$. 
If we apply a projective transformation to the discrete (semi-discrete) T-surface we obtain a  so-called discrete (semi-discrete) cone-net (cf.\ \cite{kilian}) as each vertical strip turns into a conical patch. These cone-nets are special in the way that the tips of all cones are collinear \cite[page 69]{boeklen}; therefore we call them {\it axial cone-nets}. For the illustration of a T-hedron and its projective transformation we refer to \cite[Fig.\ 5]{kilian}. 

It is well known (e.g.\ \cite{wegner}) that in general projective transformations only preserve infinitesimal flexions but not continuous ones. 
Thus we are aiming to determine the necessary and sufficient conditions for continuous flexibility of discrete (semi-discrete) axial cone-nets. 
In analogy to the notation introduced by Sauer and Graf \cite{graf} we abbreviate continuous flexible discrete cone-nets by the nomenclature C-hedra, which are eponymous for the contribution at hand structured as follows: 

In Section \ref{sec:aC_hedra} we determine the complete class of axial C-hedra by reducing it to the study of an overconstrained planar linkage (Section \ref{sec:link}).
 The obtained result can also be used to construct their semi-discrete analogs (Section \ref{sec:construct}). 
Moreover, we identify a novel subclass of axial P-nets within the determined class of discrete (semi-discrete) axial cone-nets. 
In Section \ref{sec:end} we conclude the paper with a generalization of this subclass and an outlook to future \medskip research.

But before we plunge in medias res, we list all examples of discrete and semi-discrete axial cone-nets with a continuous flexion known to the author: 
\begin{enumerate}[$\bullet$]
    \item 
    By fixing the location of the axis (i.e.\ the prism polyline degenerates into a point) for the class of discrete and semi-discrete T-surfaces, we obtain the first family of know examples. This subclass of {\it stretch rotation surfaces} is of great importance as the complete set of T-surfaces\footnote{With exception of translational T-surfaces, which are obtained if the axis goes to infinity.} can be reconstructed from it by a {\it parallelism operation} (for detail please see \cite{graf,SNRT2021,sauer,arvin}).
    \item
    A further family of known examples originate from curved origami. A huge variety of origami shapes can be produced by using curved creases.  But the question arise if a continuous folding motion exists towards the flat state that keeps the ruling layout\footnote{For origami shapes with only planar curved creases the answer was recently given in \cite{8osme}.} \cite{demain}. 
    This so-called {\it rigid-ruling folding} is the curved-crease analogue of rigid-folding. A special family of curved origami possessing a rigid-ruling folding are conic crease patterns with reflecting rule lines \cite{demain}, which constitute examples of semi-discrete axial cone-nets with a continuous flexion.

    A discretization of these crease patterns was done in \cite{mundi} under preservation of rigid-foldability, which results in a family of axial C-hedra. For the proof of this property a constructive approach was used in \cite{mundi} based on a compatible series of planar linkages. We will stress this method in the next section to obtain the complete class of axial C-hedra. 
\end{enumerate}


\section{Determination of axial C-hedra}\label{sec:aC_hedra}

For the determination of axial C-hedra 
it is sufficient to consider only three consecutive  conical strips according to \cite[Theorem 3.2]{schief} already mentioned in Section \ref{sec:intro}. The tips of the three involved discrete cones $\Lambda_k$ are denoted by $S_k$ for $k=1,2,3$ and are located on the axis $q$ of the axial cone-net, which is assumed to be vertical\footnote{As this is the usual setup \cite{graf,sauer} for stretch-rotation surfaces, whose relation to the obtained results is stressed latter on.}.

\begin{figure}[b]
\begin{overpic}
    [height=60mm,angle=90]{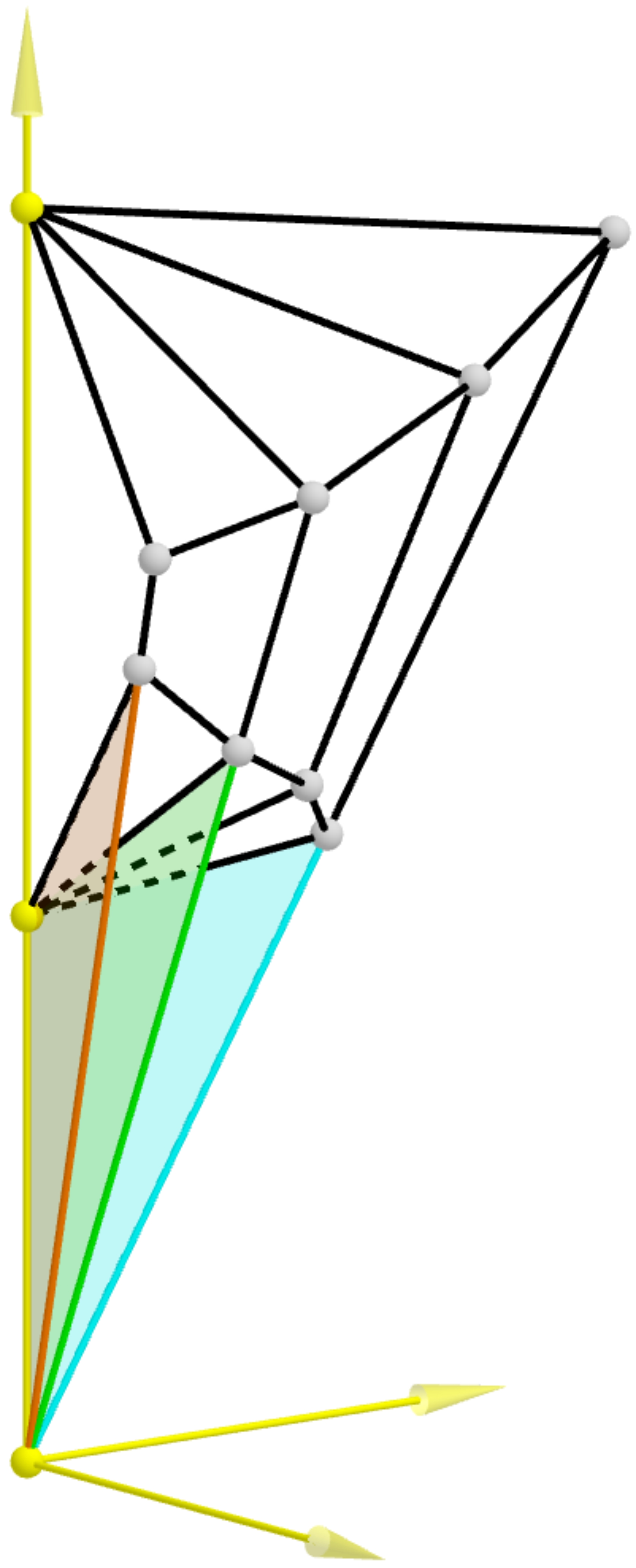}
\begin{scriptsize}
\put(13,-2){\makebox(0,0){\rotatebox{90}{$S_3$}}}
\put(35,-1){\makebox(0,0){\rotatebox{90}{$q$}}}
\put(58,-1.5){\makebox(0,0){\rotatebox{90}{$S_1$}}}
\put(93,-1.5){\makebox(0,0){\rotatebox{90}{$S_2$}}}
\put(97,25){\makebox(0,0){\rotatebox{90}{$x$}}}
\put(86,31){\makebox(0,0){\rotatebox{90}{$y$}}}
\put(3,4){\makebox(0,0){\rotatebox{90}{$z$}}}
\put(62,4){\makebox(0,0){\rotatebox{90}{$\pi_0$}}}
\put(62,8.7){\makebox(0,0){\rotatebox{90}{$\pi_1$}}}
\put(62,13.7){\makebox(0,0){\rotatebox{90}{$\pi_i$}}}
\put(42,5.5){\makebox(0,0){\rotatebox{90}{$A_0$}}}
\put(35,6.3){\makebox(0,0){\rotatebox{90}{$B_0$}}}
\put(44,13.7){\makebox(0,0){\rotatebox{90}{$A_1^*$}}}
\put(33,23.5){\makebox(0,0){\rotatebox{90}{$B_1^*$}}}
\put(55,23.7){\makebox(0,0){\rotatebox{90}{$A_i^*$}}}
\put(12,41){\makebox(0,0){\rotatebox{90}{$B_i^*$}}}
\end{scriptsize}     
  \end{overpic} 
\hfill
 \begin{overpic}
    [height=57mm,angle=90]{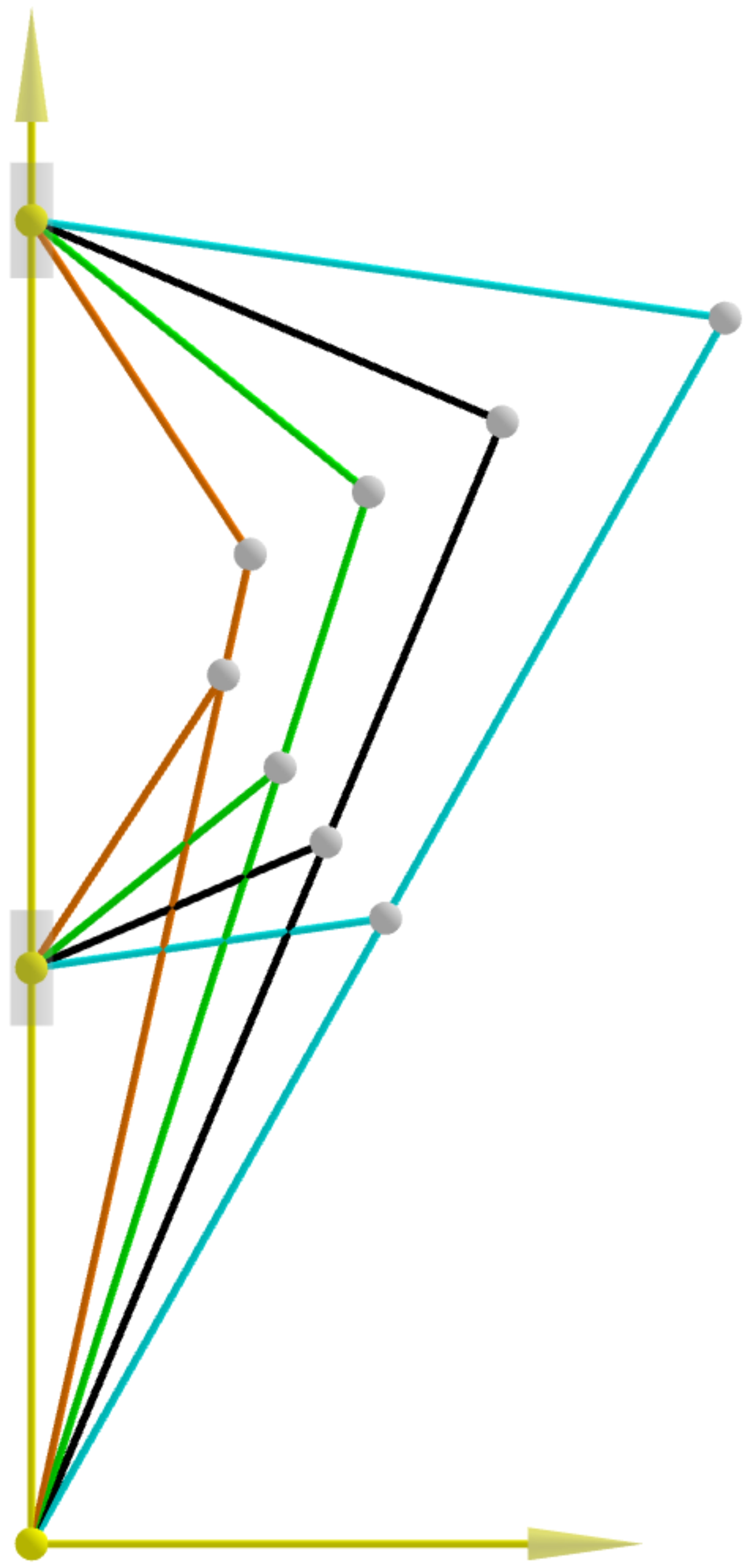}
\begin{scriptsize}
\put(14,-2){\makebox(0,0){\rotatebox{90}{$S_3$}}}
\put(37,-1){\makebox(0,0){\rotatebox{90}{$q$}}}
\put(62,-1.5){\makebox(0,0){\rotatebox{90}{$S_1$}}}
\put(97,-1.5){\makebox(0,0){\rotatebox{90}{$S_2$}}}
\put(96,37.5){\makebox(0,0){\rotatebox{90}{$x$}}}
\put(3,4){\makebox(0,0){\rotatebox{90}{$z$}}}
\put(42,10.5){\makebox(0,0){\rotatebox{90}{$A_0$}}}
\put(35,12){\makebox(0,0){\rotatebox{90}{$B_0$}}}
\put(46.5,21){\makebox(0,0){\rotatebox{90}{$A_1$}}}
\put(31.5,20.5){\makebox(0,0){\rotatebox{90}{$B_1$}}}
\put(59,28){\makebox(0,0){\rotatebox{90}{$A_i$}}}
\put(17,46){\makebox(0,0){\rotatebox{90}{$B_i$}}}
\end{scriptsize}       
  \end{overpic} 
  \medskip
\hfill
\caption{Schematic sketch of  three consecutive  conical strips of an axial C-hedra in an axonometric view (left) and the corresponding overconstrained planar linkage $\mathcal{L}$ located in the $xz$-plane (right) where the sublinkages $L_0$, $L_1$ and $L_i$ are highlighted in  orange, green and blue, respectively. In order to save space the pictures are rotated by 90 degrees.   
}
  \label{fig1}
\end{figure}
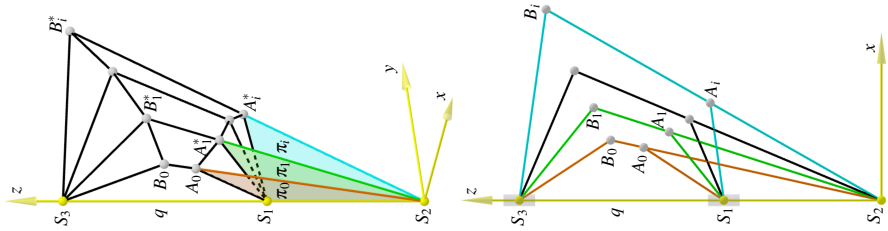

\subsection{The planar linkage $\mathcal{L}$}\label{sec:link}

We consider the cuts of the cone-net by planes $\pi_i$ (for $i=0,1,\ldots,p$ and $p\geq3$) containing the edges of $\Lambda_k$ and the axis $q$ (see Fig.\ \ref{fig1}).  
All the obtained planar cuts can then be rotated into the plane $\pi_0$.
The resulting planar structure can be interpreted as a planar linkage $\mathcal{L}$ which consists of $6+3p$ systems and $7+5p$ joints with 1-DoF. 
According to the formula of Gr\"ubler we get as degree of freedom for the linkage the value $1-p$. But this linkage has to have an overconstrained mobility as the aimed continuous flexion of the three consecutive conical strips induces a 1-DoF motion to $\mathcal{L}$. In the following we study the necessary and sufficient conditions for $\mathcal{L}$ to have an overconstrained 1-DoF motion. 

Let us start by introducing the used notation and coordinatization.
Similar to \cite{KMN1,SNRT2021,arvin} we choose our Cartesian coordinate frame in a way that the $z$-axis coincide with $q$ and that $\pi_0$ equals the $xz$-plane. Moreover, we can assume that its origin coincides with $S_2$ and that $S_1$ is located on the positive part of the $z$-axis. In addition we can orient the $x$-axis in a way that $A_0$ has a positive $x$-coordinate. This yields the following coordinatization for the points illustrated in Fig.\ \ref{fig1}:
\begin{equation}\label{eq:input}
 S_1:=(0,0,a)^T, \quad
 S_3:=(0,0,b)^T, \quad
 A_i:=(d_i,0,z_i)^T, \quad
 B_i=v_iA_i,
\end{equation}
with $a\geq 0$, $d_0\geq 0$. Further we can assume $a \neq 0$ and $b\neq 0$ 
to exclude the coincidence of adjacent cones, as the arrangement of two cones has always a trivial flexion. Moreover, we use the following notation of non-zero bar lengths:
\begin{equation}
 s_i:=\overline{S_2A_i}>0, \quad
 t_i:=\overline{S_1A_i}>0, \quad
 u_i:=\overline{S_3B_i}>0.
\end{equation}
Therefore the length of the bar $\overline{S_2B_i}>0$ is given by $|v_i|s_i$ with $v_i\neq 0$.
Now we can express the coordinates of $A_i$ in dependence of $s_i$, $t_i$ and $a$ as:
\begin{equation}
d_i= \tfrac{\sqrt{ 2a^2s_i^2 + 2a^2t_i^2 +2s_i^2t_i^2  - a^4  - s_i^4  - t_i^4}}{2a}, \quad
    z_i := \tfrac{a^2 + s_i^2 - t_i^2}{2a}.
\end{equation}
The idea is to use the slider associated with $S_1$ (cf.\ Fig.\ \ref{fig1}) as active joint; i.e.\ $a$ acts as parameter of the 1-DoF of $\mathcal{L}$. We can now compute the  position of $S_3$ in dependence of the sublinkage with index $0$ of $\mathcal{L}$, which we call the initial sublinkage $L_0$. The equation $\|B_0-S_3\|^2-u_0^2=0$ can be solved for $b$ yielding following two solutions:
\begin{equation}\label{eq:b}
    b_{\pm} := \tfrac{(a^2 + s_0^2 - t_0^2)v_0 \pm 
    \sqrt{[a^4  -2(s_0^2 + t_0^2)a^2 + (s_0 - t_0)^2(s_0 + t_0)^2]v_0^2 + 4u_0^2a^2}}{2a}. 
\end{equation}
Using this $b_{\pm}$ value we can compute $\|B_j-S_3\|^2-u_j^2=0$ for any $j\in\left\{1,\ldots, p\right\}$ yielding the expression $W_{j\pm}$.  
Next we get rid of the square root of Eq.\ (\ref{eq:b}) in $W_{j\pm}$ by an appropriate rearrangement and a subsequent squaring. 
The resulting expression\footnote{Is the same for $W_{j-}$ and $W_{j+}$.} of the form 
$f_ja^4+g_ja^2+h_j=0$
 has to be fulfilled independent of the motion parameter $a$. Therefore the coefficients $f_j,g_j,h_j$, which are given in detail in Appendix A, have to vanish. 
The solutions of the resulting system of equations $f_j=g_j=h_j=0$, obtained by a detailed discussion of cases done in Appendix A, are summarized in the following theorem:
\begin{theorem}\label{thm:1}
The linkage $\mathcal{L}$ has an overconstrained 1-DoF mobility if and only if each sublinkage with index $j$ for $j\in\left\{1,\ldots, p\right\}$ belongs to one of the following cases defined by the initial linkage $L_0$:
\begin{enumerate}[1.]
    \item Central Scaling:
    \begin{enumerate}[a.]
        \item 
        For $v_0=\tfrac{u_0}{t_0}$ and either $z_0\geq 0$ and $b_+$ or $z_0\leq 0$ and $b_-$: $v_j=\tfrac{u_0}{t_0}$, $u_j=\tfrac{u_0t_j}{t_0}$ 
        \item 
        For $v_0=-\tfrac{u_0}{t_0}$ and either $z_0\leq 0$ and $b_+$ or  $z_0\geq 0$ and $b_-$:  $v_j=-\tfrac{u_0}{t_0}$, $u_j=\tfrac{u_0t_j}{t_0}$
    \end{enumerate}
    \item Perspective Collineation: 
    \begin{enumerate}[a.]
        \item 
        For $v_0=-\tfrac{u_0}{t_0}$ and either $z_0\geq 0$ and $b_+$ or $z_0\leq 0$ and $b_-$:  \newline
        $u_j = \tfrac{\sqrt{t_0v_j(s_j^2t_0v_j + s_0^2u_0 - t_0^2u_0)}}{t_0}$, $v_j = -\tfrac{u_0(s_0^2 - t_0^2)}{t_0(s_j ^2 - t_j ^2)}$ 
        \item 
        For $v_0=\tfrac{u_0}{t_0}$ and either $z_0\leq 0$ and $b_+$ or  $z_0\geq 0$ and $b_-$: \newline  
        $u_j = \tfrac{\sqrt{t_0v_j(s_j^2t_0v_j - s_0^2u_0 + t_0^2u_0)}}{t_0}$, $v_j = \tfrac{u_0(s_0^2 - t_0^2)}{t_0(s_j ^2 - t_j ^2)}$ 
    \end{enumerate}
    \item Central Perspectivity: 
    $v_j=v_0$, $u_j = \sqrt{s_j^2v_0^2 -s_0^2v_0^2 + u_0^2}$, $t_j = \sqrt{s_j^2 -s_0^2 + t_0^2}$  
\end{enumerate}
\end{theorem}

Note that in case 3 the initial sublinkage can be arbitrary, in contrast to the cases 1 and 2, where the relation $v_0=\pm \tfrac{u_0}{t_0}$ of the {\it basic proportionality theorem} has to be fulfilled. If in case 3 this relation holds additionally true, then it also belongs to either case 1 or case 2 depending on the conditions implied by $L_0$. 
These are the only possible sublinkages belonging to two classes at the same time. Beside these linkages it is impossible that  $\mathcal{L}$ 
contains sublinkages of different classes, which can also be seen from the conditions implied by $L_0$ (cf.\ Theorem \ref{thm:1}). 

On the other hand the set of sublinkages for the index $j$ is only $1$-dimensional for case 3 but $2$-dimensional for the cases 1 and 2. This becomes more evident by the knowledge of the geometry of these cases, which is explained next.\medskip

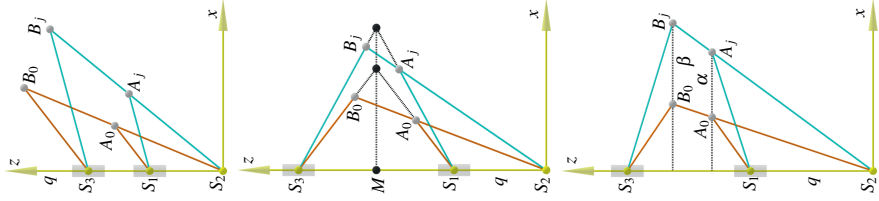
\begin{figure}[t]
\begin{overpic}
    [width=24mm,angle=90]{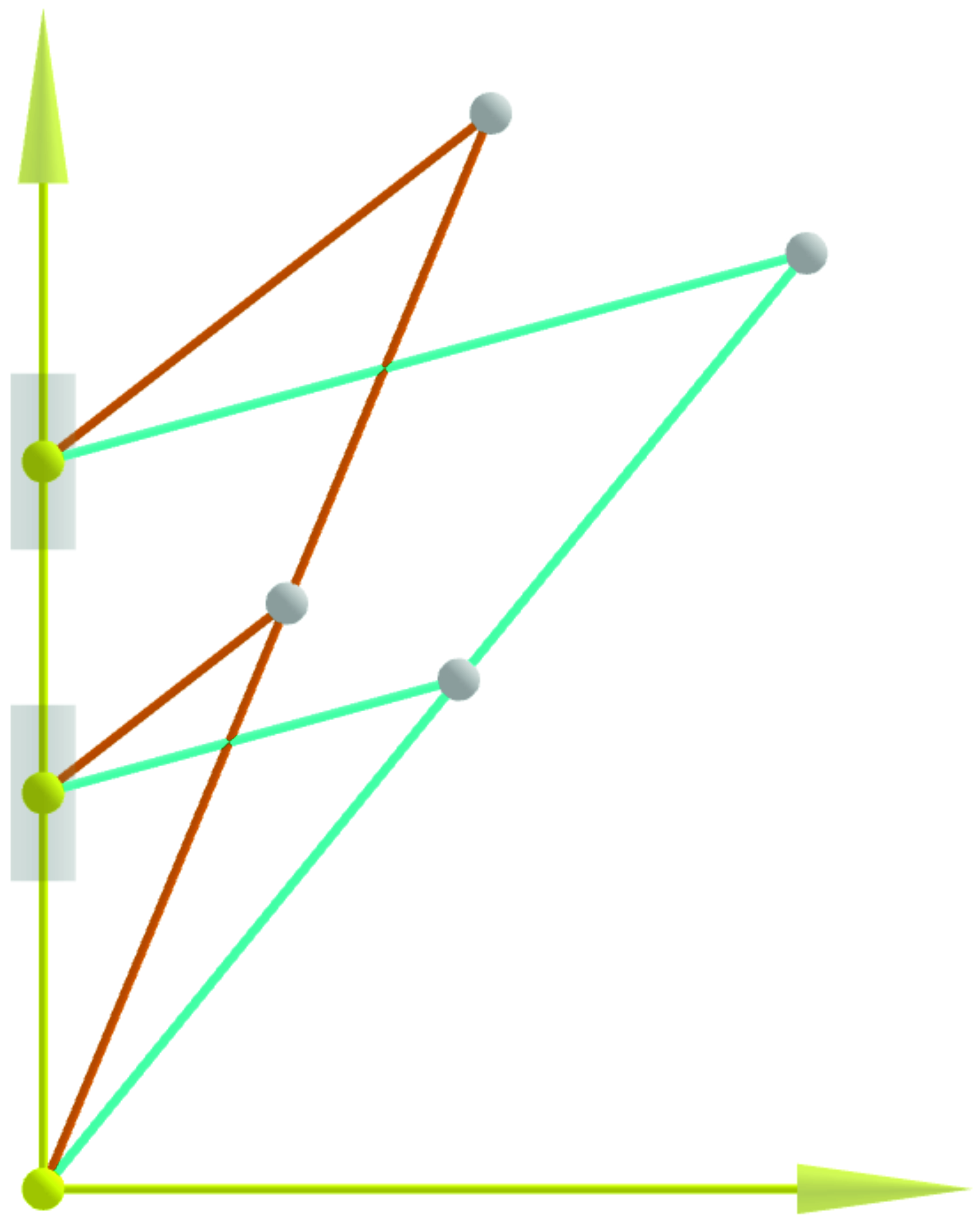}
\begin{scriptsize}
\put(37,-3){\makebox(0,0){\rotatebox{90}{$S_3$}}}
\put(21,-1){\makebox(0,0){\rotatebox{90}{$q$}}}
\put(65,-2.5){\makebox(0,0){\rotatebox{90}{$S_1$}}}
\put(95,-2){\makebox(0,0){\rotatebox{90}{$S_2$}}}
\put(93,73){\makebox(0,0){\rotatebox{90}{$x$}}}
\put(5,7){\makebox(0,0){\rotatebox{90}{$z$}}}
\put(46,17){\makebox(0,0){\rotatebox{90}{$A_0$}}}
\put(12,46){\makebox(0,0){\rotatebox{90}{$B_0$}}}
\put(59,43){\makebox(0,0){\rotatebox{90}{$A_j$}}}
\put(15,68){\makebox(0,0){\rotatebox{90}{$B_j$}}}
\end{scriptsize}        
  \end{overpic} 
\hfill
\begin{overpic}
    [width=24mm,angle=90]{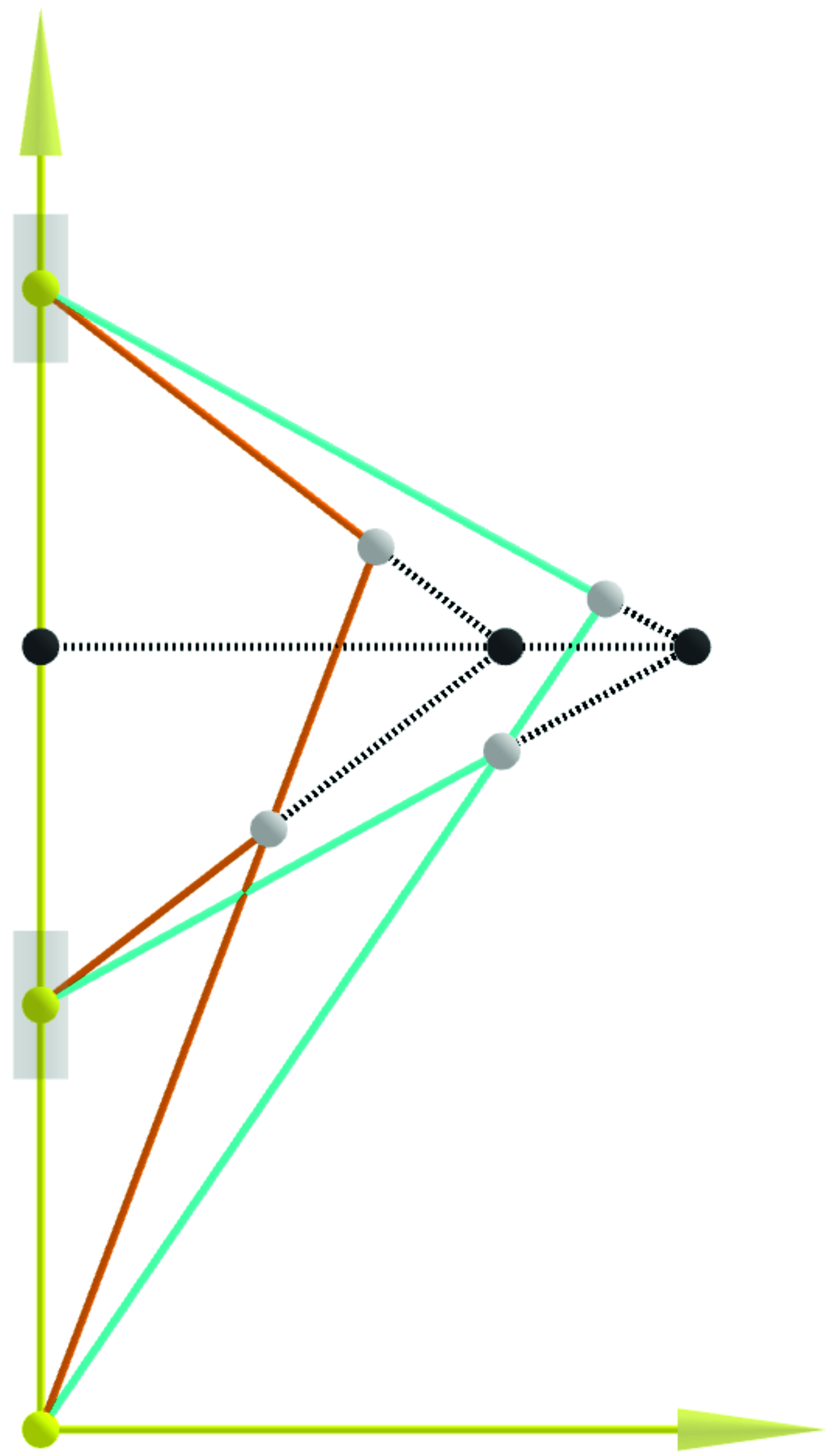}
\begin{scriptsize}
\put(45,-2){\makebox(0,0){\rotatebox{90}{$M$}}}
\put(19,-2){\makebox(0,0){\rotatebox{90}{$S_3$}}}
\put(84,-1){\makebox(0,0){\rotatebox{90}{$q$}}}
\put(68,-1.5){\makebox(0,0){\rotatebox{90}{$S_1$}}}
\put(97,-2){\makebox(0,0){\rotatebox{90}{$S_2$}}}
\put(94.5,52){\makebox(0,0){\rotatebox{90}{$x$}}}
\put(4,5.5){\makebox(0,0){\rotatebox{90}{$z$}}}
\put(53,15){\makebox(0,0){\rotatebox{90}{$A_0$}}}
\put(37.5,20.9){\makebox(0,0){\rotatebox{90}{$B_0$}}}
\put(55,38){\makebox(0,0){\rotatebox{90}{$A_j$}}}
\put(37,43){\makebox(0,0){\rotatebox{90}{$B_j$}}}
\end{scriptsize}      
  \end{overpic} 
\hfill
\begin{overpic}
    [width=24mm,angle=90]{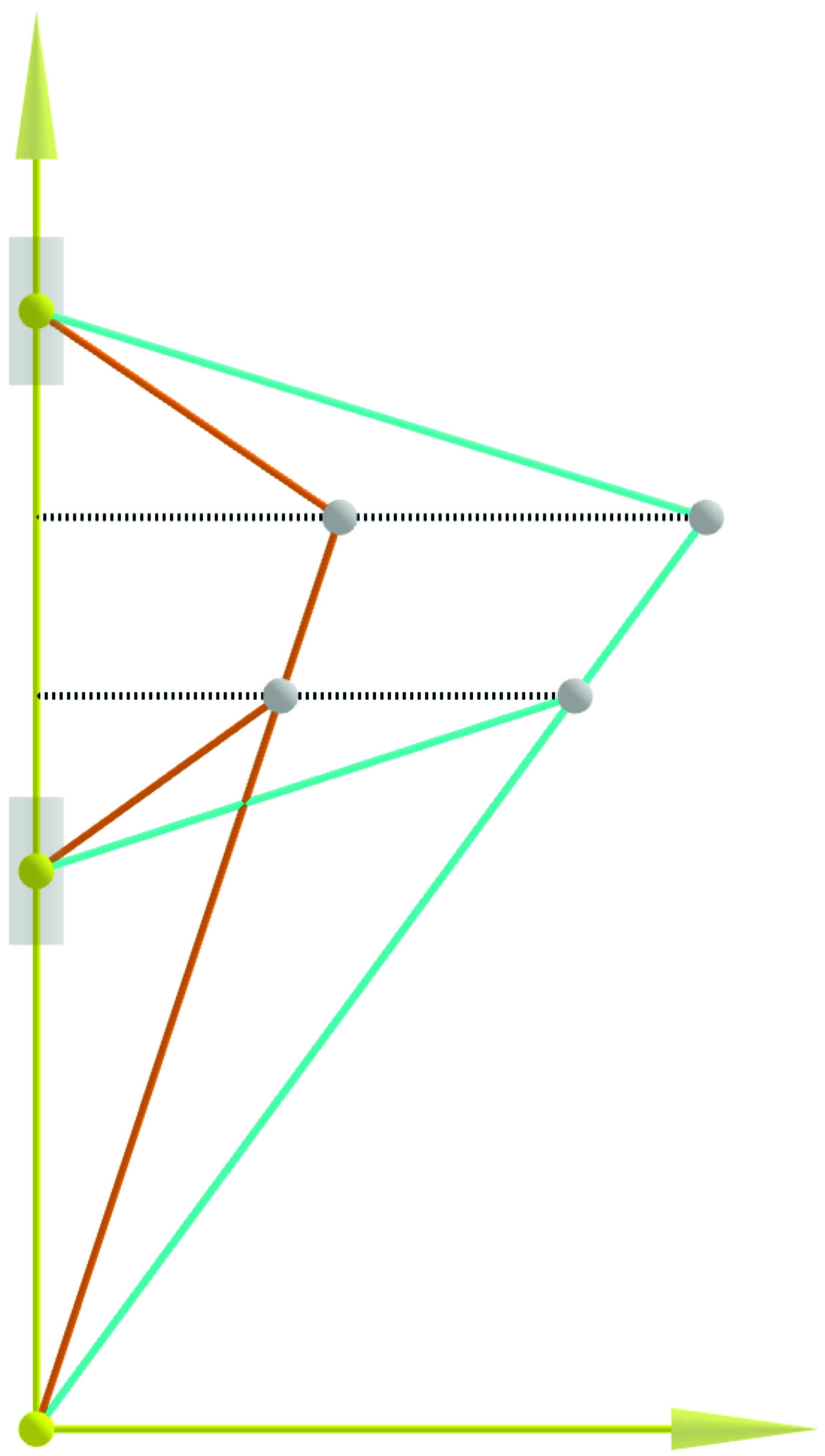}
\begin{scriptsize}
\put(21,-2){\makebox(0,0){\rotatebox{90}{$S_3$}}}
\put(80,-1){\makebox(0,0){\rotatebox{90}{$q$}}}
\put(60,-1.5){\makebox(0,0){\rotatebox{90}{$S_1$}}}
\put(97,-2){\makebox(0,0){\rotatebox{90}{$S_2$}}}
\put(94.5,52){\makebox(0,0){\rotatebox{90}{$x$}}}
\put(4,5.5){\makebox(0,0){\rotatebox{90}{$z$}}}
\put(44.5,15.5){\makebox(0,0){\rotatebox{90}{$A_0$}}}
\put(39.5,26.5){\makebox(0,0){\rotatebox{90}{$B_0$}}}
\put(53,41){\makebox(0,0){\rotatebox{90}{$A_j$}}}
\put(32,49.5){\makebox(0,0){\rotatebox{90}{$B_j$}}}
\put(39.5,37){\makebox(0,0){\rotatebox{90}{$\beta$}}}
\put(45,31){\makebox(0,0){\rotatebox{90}{$\alpha$}}}
\end{scriptsize}       
  \end{overpic} 
  \bigskip
\hfill
\caption{Illustration of the central scaling $\sigma_1$ (left), 
perspective collineation $\sigma_2$ (center) and
central perspectivity $\sigma_3$ (right), respectively. In order to save space the pictures are rotated by 90 degrees.   
}
  \label{fig2}
\end{figure}

\noindent
{\bf Geometric interpretation of Theorem \ref{thm:1}.}
Based on the expression given in Theorem \ref{thm:1} it can easily be verified by direct computations that case $k$ corresponds to the linear map $\sigma_k$ for which $\sigma_k(A_0)=B_0$ and $\sigma_k(A_j)=B_j$ hold true for $k\in\left\{1,2,3\right\}$.
\begin{enumerate}[{{ad} 1:}]
    \item
    There exists a central scaling $\sigma_1$ which maps $S_1$ to $S_3$  with center $S_2$. 
    \item
    There exists a perspective collineation $\sigma_2$ which maps $S_1$ to $S_3$ with center $S_2$ and the perpendicular to $q$ through the midpoint $M$ of $S_1$ and $S_3$ as axis.
    \item
    We consider the two lines $\alpha$ and $\beta$ orthogonal to the axis $q$ passing through the points $A_0$ and $B_0$, respectively. 
    There exists a central perspectivity $\sigma_3$ with center $S_2$ mapping points of $\alpha$ to points of $\beta$. 
\end{enumerate}
The three mappings $\sigma_1$, $\sigma_2$ and $\sigma_3$  are illustrated in in Fig.\ \ref{fig2}.

\begin{remark}
The limit cases of $\sigma_k$ where at least one cone tip goes to infinity are discussed in Appendix B. \hfill $\diamond$
\end{remark}

\subsection{Construction of continuous flexible (semi-)discrete cone-nets}\label{sec:construct}
Based on Theorem \ref{thm:1} we can construct three consecutive conical strips with a continuous flexibility. To do so, we assume that our initial linkage $L_0$ is given as an input. 
As further input we construct a polyline $A_0,A_1^*,\ldots ,A_p^*$ (cf.\ Fig.\ \ref{fig1}) with 
    \begin{equation}
        A_j^*=(d_j\cos{\phi_j},d_j\cos{\phi_j},w_k)^T \quad \text{and} \quad 
 \begin{cases}
     \text{$w_k=z_j$ for $k=1,2$,} \\
     \text{$w_k=z_0$ for $k=3$,}
 \end{cases}
    \end{equation}
    for case $k$. 
    A rotation $\delta_{-\phi_j}$ of $A_j^*$ by the angle $-\phi_j$ about the $z$-axis yields the missing input $A_j$ of Eq.\ (\ref{eq:input}) for the construction of $\mathcal{L}$. Then we rotate the sublinkage $L_j$ of $\mathcal{L}$ back by the operation $\delta_{\phi_j}$. The obtained point $B_j^*:=\delta_{\phi_j}(B_j)$ completes the $(3\times p)$ C-hedral patch. In the case $k=3$ the patch belongs to a T-hedral stretch rotation surface (cf.\ Section \ref{sec:MOR}). For $k=1,2$ and $p\geq4$ we obtain a novel family of three consecutive conical strips with a continuous flexion up to the author's knowledge\footnote{For $p=3$ the resulting $(3\times 3)$ patches have to be covered by the involved and quite abstract classification  of Izmestiev \cite{Izm17} due to its claim of completeness.}.
    \begin{remark}\label{rmk:stachel}
    If $S_2$ goes to infinity (i.e.\  $\Lambda_2$ degenerates into a cylinder), then we obtain for case 1 in the limit the translational case mentioned by Stachel \cite[Sec.\ 1.2]{stachel} and for case 2 the planar-symmetric case already known to Kokotsakis \cite[§ 18]{kokotsakis}, respectively (see \cite[Fig.\ 2]{stachel} for the illustration of both limit cases). \hfill $\diamond$
    \end{remark}

\noindent
{\bf Construction of the corresponding semi-discrete surface patches.}
We can also apply the construction principle described above to a smooth input curve instead of a discrete one. In detail we replace the polyline $A_0,A_1^*,\ldots ,A_p^*$ by a smooth curve $\Vkt a^*(r)$ with parameter $r\in\left[0;1\right]$ and $\Vkt a^*(0)=A_0$ for the respective case k; i.e.\
\begin{equation}
 \Vkt a^*(r)=(d(r)\cos{\phi(r)},d(r)\cos{\phi(r)},w_k)^T   \,\,\,\,\, \text{with} \,\,\,\,\, 
 \begin{cases}
     \text{$w_k=z(r)$ for $k=1,2$,} \\
     \text{$w_k=z_0$ for $k=3$.}
 \end{cases}
\end{equation}
Then we apply again the rotation $\delta_{-\phi(r)}$ to $\Vkt a^*(r)$ to obtain $\Vkt a(r)$. The transformation of this curve with $\sigma_k$ results in $\Vkt b(r)$. Finally, $\Vkt b(r)$ is rotate back via $\delta_{\phi(r)}$ to obtain $\Vkt b^*(r)$ completing the continuous flexible arrangement of three smooth conical strips.

\begin{remark}
    Instead of applying the three linear transformations  $\delta_{-\phi}$, $\sigma_k$ and $\delta_{\phi}$ consecutively in the discrete as well as semi-discrete setting, one could replace them directly by the spatial analog of the map $\sigma_k$ for $k\in\left\{1,2,3\right\}$. \hfill $\diamond$
\end{remark}

By a subsequent combination of initial linkages of case 1 and case 2 we can generate novel discrete (semi-discrete) axial cone-nets of arbitrary dimensions with a continuous flexion. We call this new subclass discrete (semi-discrete) {\it axial P-nets} as for both cases 1 and 2 the proportion (P) $v_0=\pm\tfrac{u_0}{t_0}$ of the {\it intercept theorem} has to hold. 
Known special cases of these axial P-nets are the  smooth and discrete conic crease patterns with reflecting rule lines (cf.\ \cite{demain,mundi}).


\section{Conclusion and future work}\label{sec:end}

By using the already mentioned parallelism operation  (cf.\ Section \ref{sec:MOR}) one can even generalize the subclass of axial P-nets. The resulting general  P-nets  constitute a rich\footnote{A simple parameter count reveals that the design space of a square patch has the same dimension as the corresponding one for T-hedra.} novel class of continuous flexible discrete (semi-discrete) surfaces, which allow direct access to their spatial shapes by three control polylines (similar to T-surfaces). 
This intuitive method makes them suitable for transformable design tasks using interactive tools like the  Rhino/Grasshopper plugin ``Scutes'' \cite{voss,SNRT2021}.
The implementation of these general P-nets to ``Scutes'' is dedicated to future research as well as the parametrization of their isometric deformation required for coding. Further we plan a detailed discussion of their flexion limits, bifurcation configurations and possible tubular arrangements (in analogy to \cite{KMN1}).

An interesting open question remains towards the existence of non-axial C-hedra and their semi-discrete analogs.

\newpage

\begin{acknowledgement}
The research is funded by project F77 (SFB ``Advanced Computational Design'') of the Austrian Science Fund FWF. Moreover the author wants to thank Christian M\"uller for providing reference \cite{boeklen}.
\end{acknowledgement}

\section*{Appendix A: Proof of Theorem \ref{thm:1}}

In the following we prove that the system of equations $f_j=g_j=h_j=0$ with
\begin{equation}
\begin{split}
f_j:=&(v_0 - v_j)[s_0^2v_jv_0^2 - u_0^2v_j + (u_j^2-s_j^2v_j^2)v_0], \\
g_j:=&
(s_0^2 + s_j^2 - t_0^2 - t_j^2)v_0s_j^2v_j^3 
+ 2(s_0^2u_0^2 - t_0^2u_j^2)v_0^2 
- (u_0 - u_j)^2(u_0 + u_j)^2 \\
&-s_j^4v_j^4 
- s_0^4v_0^4 
+ 2[(t_0^2 - s_0^2)s_j^2 + s_0^2t_j^2]v_0^2v_j^2 + 2(s_j^2u_j^2 - t_j^2u_0^2)v_j^2 \\
&+ (s_0^2v_0^2 - u_0^2 - u_j^2)(s_0^2 + s_j^2 - t_0^2 - t_j^2)v_0v_j, \\
h_j:=&
[(s_0^2 - t_0^2)v_0u_j^2 -(s_0^2 - t_0^2)v_0s_j^2v_j^2 + (s_0^2s_j^2 - s_0^2t_j^2)v_0^2v_j + (t_j^2 - s_j^2)u_0^2v_j] \\
&[(t_j^2 -s_j^2)v_j + v_0(s_0^2 - t_0^2)],
\end{split}
\end{equation}
has only the solutions listed in Theorem \ref{thm:1}. 

As  $t_j$ only appears in $g_j$ and $h_j$ we eliminate it from these two expressions by means of resultant; i.e.\ $m_j:=Res(h_j,g_j,t_j)$. In a further step we compute $Res(f_j,m_j,v_j)$ which factors into $v_0^8u_j^8s_j^{16}F_-^{16}F_+^{16}G_-^{6}G_+^{6}H^{16}$  with
\begin{equation}\label{eq:factors}
   F_{\mp}=s_0v_0 \mp u_0, \quad
   G_{\mp}=t_0v_0 \mp u_0, \quad
   H=s_0^2v_0^2 - s_j^2v_0^2 - u_0^2 + u_j^2.
\end{equation}
As the first three factors cannot vanish, we remain with the discussion of the factors given in Eq.\ (\ref{eq:factors}). 

\begin{enumerate}[$\bullet$]
    \item $F_{\mp}=0$: From this equation we can express $v_0=\pm \tfrac{u_0}{s_0}$. Now the greatest common divisor (gcd) of $f_j$ and $m_j$ equals $u_0(s_jv_j - u_j)(s_jv_j + u_j)$. This expression cannot vanish without contradiction as the latter two factors imply $b=0$. 
    \item $G_-=0$, $F_{\mp}\neq 0$: From this equation we can express $v_0= \tfrac{u_0}{t_0}$. Then the gcd of $f_j$ and $m_j$ equals 
    $u_0(t_0v_j - u_0)(s_0^2u_0v_j - s_j^2t_0v_j^2 - t_0^2u_0v_j + t_0u_j^2)$. Thus we have to distinguish the following two cases:
    \begin{enumerate}[$\star$]
        \item $v_j = \tfrac{u_0}{t_0}$: Then the gcd of $g_j$ and $h_j$ equals $(t_0u_j + t_ju_0)(t_0u_j - t_ju_0)$. The first factor cannot vanish as all bar lengths are greater than zero. It can be seen that the second factor yields solution 1a after back substitution into $W_{j\pm}$.
        \item $u_j = \tfrac{\sqrt{t_0v_j(s_j^2t_0v_j - s_0^2u_0 + t_0^2u_0)}}{t_0}$: Now the gcd of $g_j$ and $h_j$ equals $u_0^2(s_0 - t_0)(s_0 + t_0)(t_0t_j^2v_j  - s_j^2t_0v_j + s_0^2u_0 - t_0^2u_0)$. The vanishing of the second and third factor yields $F_{\mp}= 0$, a contradiction. It can be seen that the last factor yields solution 2b after back substitution into $W_{j\pm}$.
    \end{enumerate}
    \item $G_+=0$, $F_{\mp}\neq 0$: From this equation we can express $v_0= -\tfrac{u_0}{t_0}$. Then the gcd of $f_j$ and $m_j$ equals 
    $u_0(t_0v_j + u_0)(s_0^2u_0v_j + s_j^2t_0v_j^2 - t_0^2u_0v_j - t_0u_j^2)$. Thus we have to distinguish the following two cases:
    \begin{enumerate}[$\star$]
    \item $v_j = -\tfrac{u_0}{t_0}$: Then the gcd of $g_j$ and $h_j$ equals $(t_0u_j + t_ju_0)(t_0u_j - t_ju_0)$. The first factor cannot vanish as all bar lengths are greater than zero. It can be seen that the second factor yields solution 1b after back substitution into $W_{j\pm}$.
    \item $u_j = \tfrac{\sqrt{t_0v_j(s_j^2t_0v_j + s_0^2u_0 - t_0^2u_0)}}{t_0}$: Now the gcd of $g_j$ and $h_j$ equals $u_0^2(s_0 - t_0)(s_0 + t_0)(s_j^2t_0v_j - t_0t_j^2v_j + s_0^2u_0 - t_0^2u_0)$. The vanishing of the second and third factor yields $F_{\mp}= 0$, a contradiction. It can be seen that the last factor yields solution 2a after back substitution into $W_{j\pm}$.
    \end{enumerate}
    \item 
    $H=0$, $F_{\mp}G_{\mp}\neq 0$: From this equation we can express $u_j = \sqrt{s_j^2v_0^2 - s_0^2v_0^2  + u_0^2}$. Now the gcd of $f_j$ and $m_j$ implies $v_j=v_0$. Then the gcd of $g_j$ and $h_j$ equals $v_0^2(s_0v_j - u_0)(s_0v_j + u_0)(s_0^2 - s_j^2 - t_0^2 + t_j^2)^2$. The vanishing of the second and third factor yields $F_{\mp}= 0$, a contradiction. The last factor implies solution 3. This finishes the discussion of all cases.
\end{enumerate}

\section*{Appendix B:  Discussion of limit cases}

The maps $\sigma_k$ of Section \ref{sec:link} are also well defined in the limit where $S_2$ moves to infinity. Then the central scaling degenerates in a translation and the perspective collineation into a reflection (cf.\ Remark \ref{rmk:stachel}). 
We do not have to care about cases where either 
$S_2$ and $S_1$ or $S_2$ and $S_3$ are ideal points at the same time as in this case two adjacent cones coincide yielding a trivial flexion. Therefore we only have to consider the case where $S_2$ is a finite point and $S_1$ or/and $S_3$ is/are ideal point/s. 
\begin{enumerate}[{${\sigma}_1$:}]
    \item If only $S_1$ (or $S_3$) is an ideal point then all $A_i$ (or all $B_i$) are located at the line at infinity, thus the cone $\Lambda_1$ (or $\Lambda_3$) degenerates into the ideal plane, which is not feasible for the construction of a real PQ-mesh. However, we get a realizable configuration  if both  $S_1$ and $S_3$ are ideal points. In this case $\sigma_1$ has to be defined by $A_0\mapsto B_0$, as $S_1=S_3$ remains fixed under this mapping.  
    \item In this case both points $S_1$ and $S_3$ have to be ideal points, but then we end up exactly with the same case discussed before.
    \item This map is also well-defined if $S_1$ or/and $S_3$ is/are ideal point/s. 
\end{enumerate}

\begin{remark}
Note that these limits cover all parabolic cases mentioned in \cite{demain,mundi}. \hfill $\diamond$
\end{remark}
  
\end{document}